\pdfoutput=1

\documentclass[11pt]{article}

\usepackage{acl}

\usepackage{times}
\usepackage{latexsym}
\usepackage{graphicx}
\usepackage{enumitem}

\usepackage{enumitem}
\usepackage{amsmath,amssymb,amsfonts}
\usepackage{bbm,dsfont}
\usepackage{mathrsfs}
\usepackage{graphicx}
\usepackage{multirow}
\usepackage{subfigure}
\usepackage{xspace}
\usepackage{booktabs}
\usepackage{comment}
\usepackage{arydshln} 
\usepackage{bbding}
\usepackage[normalem]{ulem}
\usepackage[T1]{fontenc}

\usepackage[utf8]{inputenc}

\usepackage{microtype}

\usepackage[misc]{ifsym}

\newcount\DraftStatus  
\usepackage{color}
\definecolor{darkgreen}{rgb}{0,0.5,0}
\definecolor{purple}{rgb}{1,0,1}
\definecolor{todocolor}{rgb}{0.9,0.1,0.1}
\definecolor{hycolor}{rgb}{0.7,0.7,0.3}
\definecolor{fixcolor}{rgb}{0.1,0.7,0.3}

\newcommand{\draftnote}[2]{\ifnum\DraftStatus=1
	\marginpar{
		\tiny\raggedright
		\hbadness=10000
		\def\baselinestretch{0.8}
		\textcolor{#1}{\textsf{\hspace{0pt}#2}}}
	\fi}

\newcommand{\ourapproach}{\textsc{Code-MVP}\xspace}

\makeatletter
\def\thanks#1{\protected@xdef\@thanks{\@thanks
        \protect\footnotetext{#1}}}
\makeatother

\title{\ourapproach: Multi-View Contrastive Pre-training with Program Analysis}
\title{\ourapproach: Multi-View Contrastive Pre-Training for \\Code Representation}
\title{\ourapproach: Learning to Represent Source Code from Multiple Views with Contrastive Pre-Training}

\author{
Xin Wang$^1$\textsuperscript{$\diamond$}\thanks{$\diamond$ Work conducted during an internship at Huawei Noah's Ark Lab.} \ \
Yasheng Wang$^2$ \ \
Yao Wan$^3$ \ \
Jiawei Wang$^4$ \\
{\bf
Pingyi Zhou$^{2}$ \ \
Li Li$^4$ \ \
Hao Wu$^5$ \ \
Jin Liu$^1$\textsuperscript{\Letter}\thanks{\Letter\ Corresponding author.}
}
\\
$^1$School of Computer Science, Wuhan University, China \quad
$^2$Huawei Noah’s Ark Lab\\
$^3$School of Computer Sci. \& Tech., Huazhong University of Science and Technology, China\\
$^4$Faculty of Information Technology, Monash University, Australia \\
$^5$School of Information Science and
Engineering, Yunnan University, China
\\
\texttt{\{xinwang0920, jinliu\}@whu.edu.cn}
}

\begin{document}
\maketitle
\begin{abstract}
Recent years have witnessed increasing interest in code representation learning, which aims to represent the semantics of source code into distributed vectors. Currently, various works have been proposed to represent the complex semantics of source code from different views, including plain text, Abstract Syntax Tree (AST), and several kinds of code graphs (e.g., Control/Data Flow Graph). However, most of them only consider a single view of source code independently, ignoring the correspondences among different views. In this paper, we propose to integrate different views with the natural-language description of source code into a unified framework with \textbf{M}ulti-\textbf{V}iew contrastive \textbf{P}re-training, and name our model as \ourapproach. Specifically, we first extract multiple code views using compiler tools, and learn the complementary information among them under a contrastive learning framework. Inspired by the type checking in compilation, we also design a fine-grained type inference objective in the pre-training. Experiments on three downstream tasks over five datasets demonstrate the superiority of \ourapproach when compared with several state-of-the-art baselines. For example, we achieve 2.4/2.3/1.1 gain in terms of MRR/MAP/Accuracy metrics on natural language code retrieval, code similarity, and code defect detection tasks, respectively.
\end{abstract}

\section{Introduction}
Code intelligence that utilizes machine learning techniques to promote the productivity of software developers, has attracted increasing interest in both communities of software engineering and artificial intelligence~\cite{Lu2021CodeXGLUEAM, Feng2020CodeBERTAP, wang2022,wan2022naturalcc, wu2021}.
To achieve code intelligence, one fundamental task is code representation learning (also known as code embedding), which aims to preserve the semantics of source code in distributed vectors~\cite{AlonZLY19}.
It can support various downstream tasks about code intelligence, including code defect detection~\cite{OmriS20, ZhaoXZTY21, Zhao0Y00021}, code summarization~\cite{wan2018improving}, code retrieval~\cite{Wan2019MultimodalAN}, and code clone detection~\cite{White2016DeepLC}.

\begin{table}[!t]
	\centering
    \resizebox{0.48\textwidth}{!}{
	    \begin{tabular}{l c c c c}
		    \toprule
	        Models & Tokens &AST & Graph& PT \\
	        \midrule
	        CodeBERT~\cite{Feng2020CodeBERTAP}&\Checkmark&\XSolidBrush&\XSolidBrush&\XSolidBrush\\
	        GraphCodeBERT~\cite{Guo2021GraphCodeBERTPC}&\Checkmark&\XSolidBrush&\Checkmark&\XSolidBrush\\
	        SynCoBERT~\cite{Wang2021SynCoBERTSM}&\Checkmark&\Checkmark&\XSolidBrush&\XSolidBrush\\
	        CodeGPT~\cite{Lu2021CodeXGLUEAM}&\Checkmark&\XSolidBrush&\XSolidBrush&\XSolidBrush\\
	        PLBART~\cite{ahmad2021unified}&\Checkmark&\XSolidBrush&\XSolidBrush&\XSolidBrush\\
	        TreeBERT~\cite{Jiang2021TreeBERTAT}&\Checkmark&\Checkmark&\XSolidBrush&\XSolidBrush\\
	        ContraCode~\cite{Phan2021CoTexTML}&\Checkmark&\XSolidBrush&\XSolidBrush&\Checkmark\\
	        CoTexT~\cite{Phan2021CoTexTML}&\Checkmark&\XSolidBrush&\XSolidBrush&\XSolidBrush\\
	        CodeT5~\cite{Wang2021CodeT5IU}&\Checkmark&\XSolidBrush&\XSolidBrush&\XSolidBrush\\
	        \hline
	        \ourapproach (Our work)&\Checkmark&\Checkmark&\Checkmark&\Checkmark\\
	       
			\bottomrule
		\end{tabular}
	}
	\caption{Comparison with current pre-trained code models. PT: Program Transformation.} 
    \label{table:intros}
\end{table}
Current approaches to code representation borrow ideas from the successful deep learning methods in natural language processing, mainly attributed to the \textit{naturalness hypothesis} in source code~\cite{allamanis2018survey}.
From our investigation, existing approaches mainly represent the source code from different views of code, including code token in plain text~\cite{iyer2016summarizing}, Abstract Syntax Tree (AST)~\cite{Bui2021InferCodeSL}, and Control/Data Flow Graphs (CFGs/DFGs) of code~\cite{Cummins2020ProGraMLGD, Wang2020BlendedPS}. 
Recently, many attempts have been made to pre-train a masked language model for source code, such as CodeBERT~\cite{Feng2020CodeBERTAP}, 
GraphCodeBERT~\cite{Guo2021GraphCodeBERTPC},
SynCoBERT~\cite{Wang2021SynCoBERTSM},
CodeGPT~\cite{Lu2021CodeXGLUEAM},
PLBART~\cite{ahmad2021unified}, 
CoTexT~\cite{Phan2021CoTexTML},
and CodeT5~\cite{Wang2021CodeT5IU}.
Table~\ref{table:intros} shows the contribution of our work when compared with current pre-trained language models for source code.

Despite much progress in code representation learning, most of them only consider a single view of source code independently, ignoring the consistency among different views~\cite{Feng2020CodeBERTAP, Lu2021CodeXGLUEAM, ahmad2021unified, Wang2021CodeT5IU}. Usually, a program, accompanied by a corresponding natural-language comment (NL), can be parsed into multiple views, e.g., the source code tokens, AST, and CFG.
We argue that these different views contain complementary semantics of the program.
For example, the source code tokens (e.g., method name identifiers) and natural-language comments always reveal the \textit{lexical} semantics of code, while the intermediate structures of code (e.g., AST and CFG) always reveal the \textit{syntactic} and \textit{executive} information of code. 
In addition, a program can also be transformed (or rewritten) into different variants that have equivalent functionality.
We think that different variants of the same program reveal the \textit{functional} information of code. 
That is, those different program variants with the same functionality are expected to represent the same semantics.

Inspired by the aforementioned insights, this paper proposes a novel \ourapproach for code representation, which aims to integrate multiple views of the code into a unified framework with multi-view contrastive pre-training. 
Concretely, we first extract multiple views of code using several compiler tools, and learn the complementary information among them under a multi-view contrastive learning framework. Meanwhile, inspired by the type checking in compilation process, we also introduce fine-grained type inference as an auxiliary task in the pre-training process to encourage the model to learn more fine-grained type information.

To summarize, the contributions of this paper are two-fold:
(1) 
We are the first to represent source code from multiple views, including the code tokens, AST, CFG, and various program equivalents, under a unified multi-view contrastive pre-training framework.  
Meanwhile, we also introduce an auxiliary task of inferring type annotations for variables.
(2) We extensively evaluate \ourapproach on three program comprehension tasks. Experimental results demonstrate the superiority of \ourapproach when compared with several state-of-the-art baselines. Specifically, \ourapproach achieves 2.4/2.3/1.1 gain on MRR/MAP/Accuracy metrics in natural language code retrieval, code similarity, and code defect detection tasks, respectively.

\section{Multiple Views of Code}
\begin{figure}[t!]
	\centering
	\includegraphics[width=0.48\textwidth]{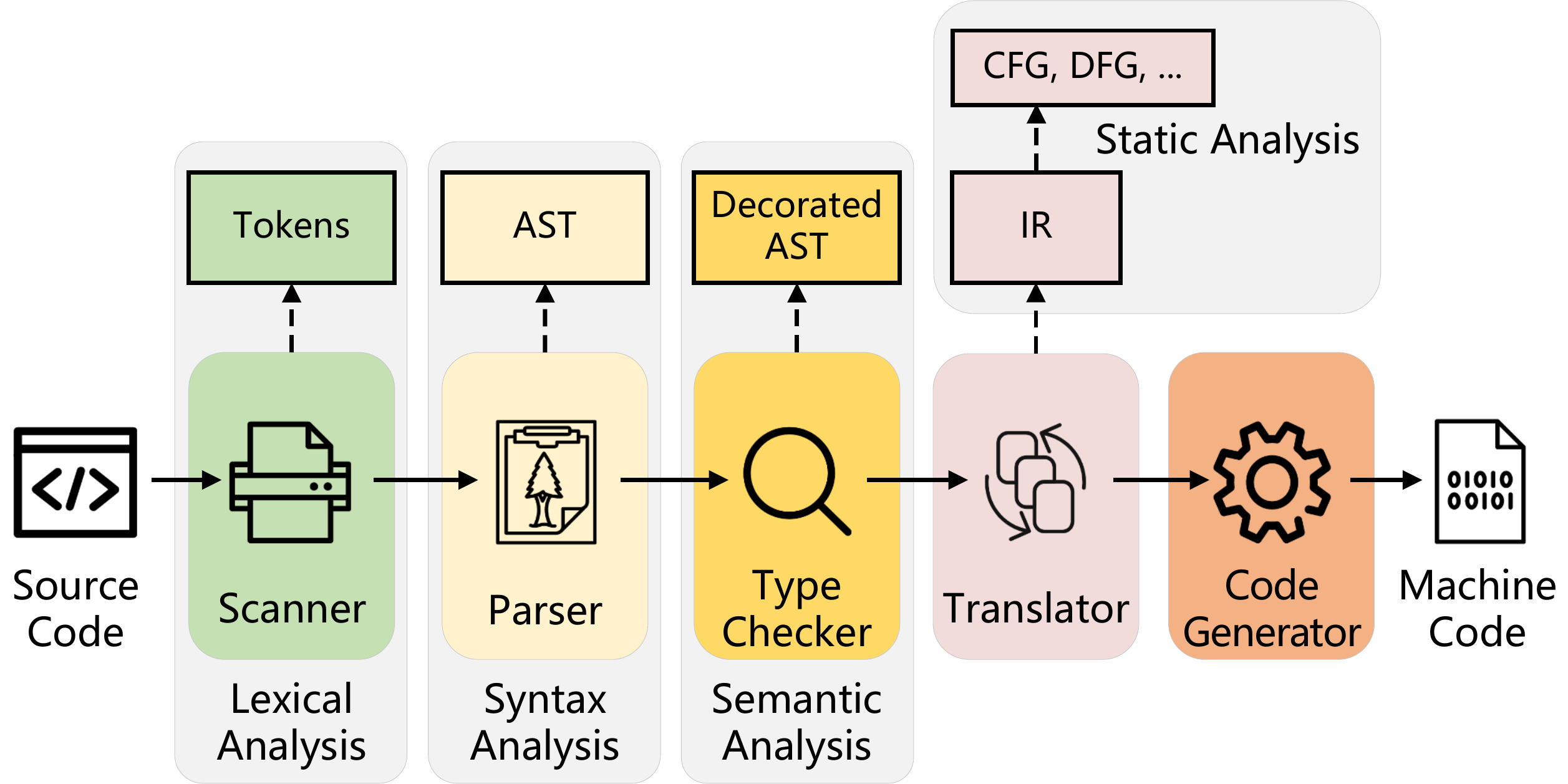}
	\caption{
    An example of converting a program from source code into machine code in compilation process.
	}
	\label{fig:pa}
\end{figure}

\begin{figure*}[!t]
	\centering
	\includegraphics[width=0.98\textwidth]{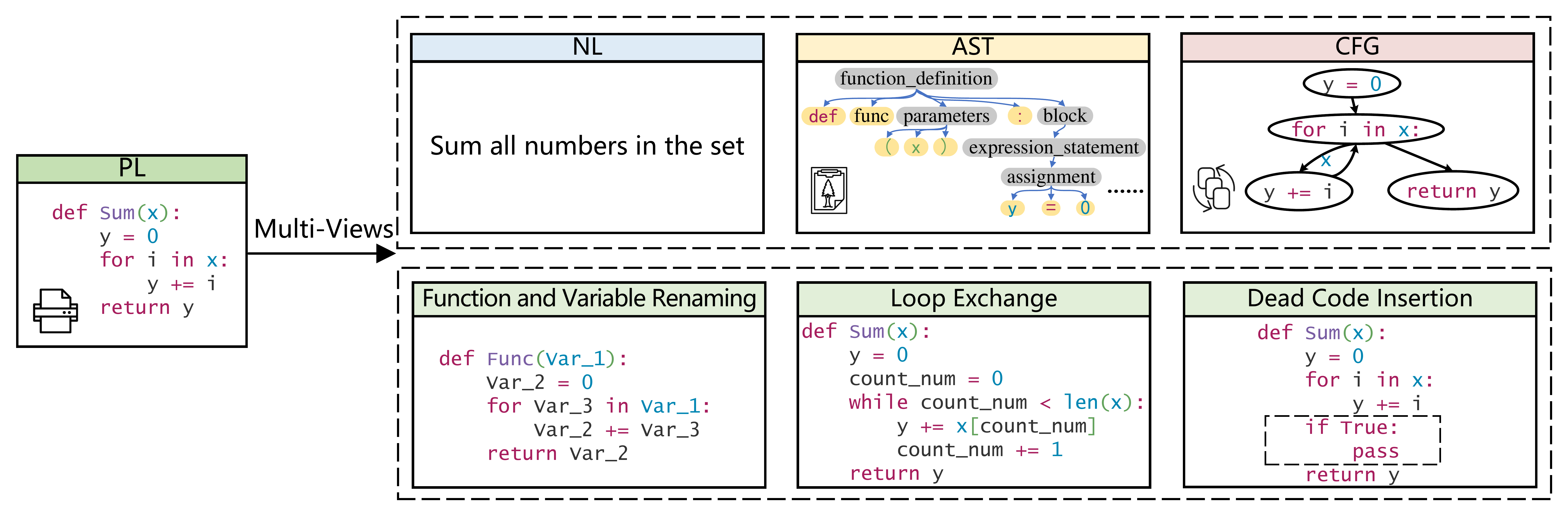}
	\caption{\label{fig:views}
    Multiple views of source code.
	}
\end{figure*}

We borrow ideas from the way that computers process the source code in compilation, where a program would be converted into multiple views.
Figure~\ref{fig:pa} shows the process of converting a program from source code to machine code. 
During this process, the compiler would automatically utilize some program analysis techniques to verify the correctness of source code, including lexical, syntax, and semantic analyses.
In the lexical analysis, a program is treated as a sequence of tokens and checked for spelling problems. In the syntax analysis, syntactic rules of programs are defined by the context-free grammar~\cite{Javed2004ContextfreeGI}. Then the program could be parsed as an AST, based on which many program transformation heuristics can be applied to rewrite the program while maintaining the same desired functionality. In the semantic analysis, semantic rules of the program are defined by the attribute grammar~\cite{Paakki1995AttributeGP}. Then the compiler could check the types of code tokens, and a decorated AST could be obtained. After the three stages above, a translator will convert the source code to its Intermediate Representation (IR), which is then considered as the basis for building Control/Data Flow Graphs (CFGs/DFGs) for further optimizations in the static analysis.
Finally, the IR of the source code should be converted into machine code to execute through a code generator. Next, we introduce how we extract different views of the source code. Figure~\ref{fig:views} illustrates multiple views of source code with an example.

\paragraph{Abstract Syntax Tree (AST).}
An AST, which is composed of leaf nodes, non-leaf nodes and edges between them, contains rich syntactic structural information of source code. In the AST, an assignment statement \texttt{y = 0} can  be  represented  by a non-leaf node \texttt{assignment} that points to three leaf nodes (\texttt{0}, \texttt{y}, and \texttt{=}).
In this paper, we parse a snippet of source code into an AST using a standard compiler tool \texttt{tree-sitter}.\footnote{https://github.com/tree-sitter/tree-sitter}.
To feed an AST into our model, 
we apply depth-first traversal to convert it into a sequence of AST tokens~\cite{Kim2021CodePB}. 

\paragraph{Control Flow Graph (CFG).}
CFG, which represents the execution semantics of the program in the form of a graph, is one intermediate representation of programs.
A CFG consists of basic blocks and directed edges between them, where each directed edge reflects the execution order of the two basic blocks in the program. We can easily traverse the CFG along directed edges to parse it into a token sequence, which reveals the execution semantics of the program.
In this paper, we use a static analyzer \texttt{Scalpel}\footnote{https://github.com/SMAT-Lab/Scalpel}~\cite{li2022} to construct the CFGs for Python code snippets.

\paragraph{Program Transformation (PT).} 
The program transformation operations aim to produce multiple variants for a given program that satisfy the same desired functionality~\cite{Rabin2020OnTG}. These different variants of a program can
help the model capture functional semantics. 
In this work, we employ the following program transformation heuristics on ASTs and rewrite one program into another equivalent variant. 

\begin{itemize}[leftmargin=*]
    \item \textbf{Function and Variable Renaming.} We randomly take new names from a set of candidates, such as \texttt{VAR\_i}, \texttt{FUNC\_i}, to
    rename the names of variables and functions in a program. This heuristic will not change the AST structure of the program, except for the textual appearance of variable and function names in the AST.
    
    \item \textbf{Loop Exchange.} The \texttt{for} and \texttt{while} loops represent the same functionality in a program. We traverse the AST to identify the \texttt{for} and \texttt{while} loop nodes, and replace \texttt{for} loops with \texttt{while} loops or vice versa. We also modify the initialization, condition and afterthought simultaneously.
    
    \item \textbf{Dead Code Insertion.} We first traverse the AST to identify several basic blocks~\cite{MendisRAC19}, and then randomly select a basic block and insert dead code snippets into it. Note that the dead code snippets are predefined and selected from a set of candidates.
\end{itemize}

\section{\ourapproach}
\subsection{Tasks and Notations}
We define the set of program samples in multiple views (i.e. NL, PL, AST, CFG, PT) as $S = \{S^1, \ldots, S^{m} \}$, where $m$ represents the number of views, 
$s_i^a\in S^{a}$ represents a program in the view of $a$. 
Given a program, the PL view denotes its textual appearance, the NL view denotes its corresponding natural-language comment, and the PT denotes the variants of this program based on program transformation.
The AST and CFG are extracted from a program using several compiler tools.
\ourapproach adopts two forms of input, i.e., single-view input $ x_i^a=\{{\texttt{<CLS>}}, s_i^a \}$ and dual-view input $x_i^{ab}=\{{\texttt{<CLS>}}, s_i^a,{\texttt{<SEP>}}, s_i^b\}$, where $a$ and $b$ denote two different views of the program. Following~\cite{devlin2018bert}, a special token $\texttt{<CLS>}$ is appended at the beginning of each input sequence, and $\texttt{<SEP>}$ is used to concatenate two sequences. 
Subsequently, the representation of $\texttt{<CLS>}$ is used to represent the entire sequence, and $\texttt{<SEP>}$ is used to split two views of sub-sequences.
Given a set of programs with their corresponding multiple views, we aim to learn the code representation by utilizing the mutual information existing in different views. 
Our intuition is to learn complementary information from multiple views of code by pulling the code under different views together and pushing the dissimilar ones apart.

\subsection{Framework Overview}
\begin{figure}[!t]
	\centering
	\includegraphics[width=0.48\textwidth]{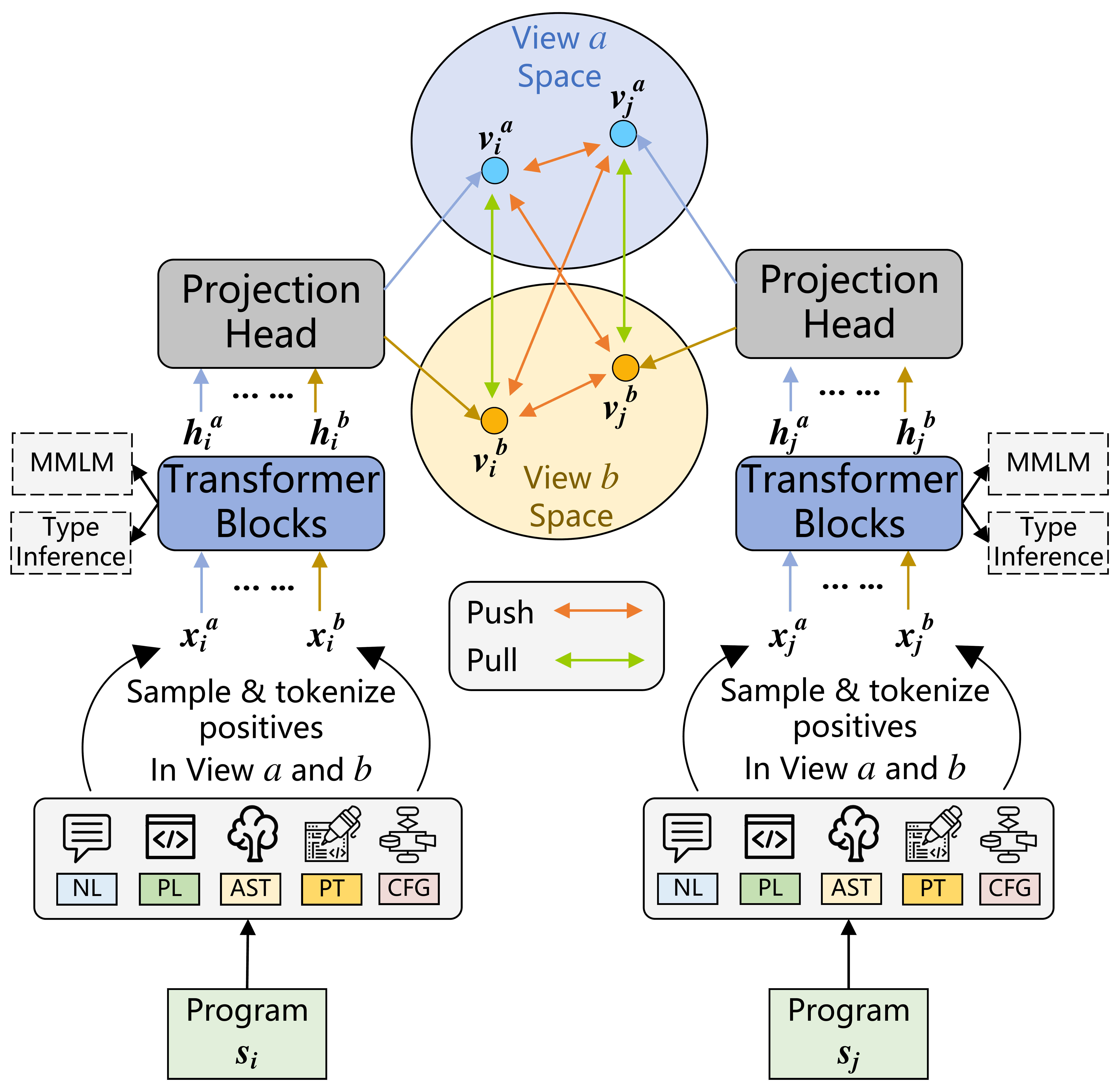}
	\caption{\label{fig:model}
		An illustration of our proposed multi-view contrastive pre-training framework.
	}
\end{figure}

Figure~\ref{fig:model} shows a simple example of our multi-view contrastive pre-training framework. Given a program $s_i$, we use the same program to construct a pair of positive samples ($x_i^a = \{{\texttt{<CLS>}}, s_i^a\}$ \textit{vs} $x_i^b = \{{\texttt{<CLS>}}, s_i^b\}$) in the form of views $a$ and $b$, as described above. We take $x_i^a$ and $x_i^b$ as the input of \ourapproach respectively. The last hidden representations of \texttt{<CLS>} tokens in the two inputs can be formulated as $\boldsymbol{h}_i^a = \ourapproach(x_i^a)$ and $\boldsymbol{h}_i^b = \ourapproach(x_i^b)$. We utilize a projection head (a two-layer MLP) to map hidden representations to a space, i.e., $\boldsymbol{v}_i^a = f(\boldsymbol{h}_i^a)$, $\boldsymbol{v}_i^b = f(\boldsymbol{h}_i^b)$. Then the multi-view contrastive objective can be performed. During the pre-training process, we also design other two pre-training tasks, i.e., fined-grained type inference (FGTI) task and multi-view masked language modeling (MMLM). 

\subsection{Multi-View Contrastive Learning}
We train \ourapproach with \textit{paired data} and \textit{unpaired data}. Paired data refers to those program samples with paired NL, while unpaired data stands for those isolated program samples without paired NL. 
Next, we explain how we construct positive and negative samples for these two cases.

\paragraph{Multi-View Positive Sampling.}
We design \textit{Single-View} (for paired and unpaired data) and \textit{Dual-View} (for paired data only, which needs the NL) methods to construct multi-view positive samples for the MVCL objective:

\begin{itemize}[leftmargin=*]
	\item \textbf{Single-View.} To bridge the gap between different views of a same program, we consider the view of a program $x_i^a$ as a positive sample w.r.t another view $x_i^b$. That is, ($x_i^a = \{{\texttt{<CLS>}}, s_i^a\}$ \textit{vs} $x_i^b = \{{\texttt{<CLS>}}, s_i^b\}$) forms an inter-view positive pair, since $x_i^a$ and $x_i^b$ are two different views of a same program $x_i$.
	\item \textbf{Dual-View.} There are a total of $C_{m}^{2}$ combinations for two views of a same program. For efficiency, we focus on the features of the program itself, and propose the \textit{NL}-conditional dual-view contrastive pre-training strategy, freezing the position of NL. Concretely, we construct a \textit{NL}-conditional inter-view positive pair by replacing the second view in the input $\{{\texttt{<CLS>}}, s_i^{\rm NL},{\texttt{<SEP>}}, s_i^a\}$ to be $\{{\texttt{<CLS>}}, s_i^{\rm NL},{\texttt{<SEP>}}, s_i^b\}$, where $\forall {a,b \neq {\rm NL}}$.
\end{itemize}
It is worth mentioning that there are many combinations to construct positive pairs. 
Some combinations are not considered in this work, such as the AST \textit{vs} PT of the same program, and the CFG \textit{vs} PT of the same program.
Simultaneously, for training efficiency and downstream applications, we comprehensively consider eight combinations. They are (1) single-view: (NL \textit{vs} PL), (NL \textit{vs} PT), (PL \textit{vs} AST), (PL \textit{vs} CFG), and (PL \textit{vs} PT); and (2) dual-view: (NL-PL \textit{vs} NL-AST), (NL-PL \textit{vs} NL-CFG), and (NL-PL \textit{vs} NL-PT).
\paragraph{Multi-View Negative Sampling.}
Since the processes of unpaired data and paired data are similar, here we take the unpaired data as an example.
We leverage \textit{in mini-batch} and \textit{cross mini-batch} sampling strategies~\cite{Chen2020ASF} to construct intra-view and inter-view negative samples, respectively. Given a mini-batch of training data $b_1 = [ x_1^a, \ldots, x_n^a ]$ in the view of $a$ with size $n$, we can easily get another positive mini-batch data $b_2 = [ {x_1^b},\ldots, {x_n^b}]$ in the view of $b$, where ($x^a_i$ \textit{vs} $x_i^b$) denotes an inter-view positive pair. For $x_i^a$, the intra-view negative samples are $\{x_j^a\}, \forall {i \neq j}$, and the inter-view negative samples are $\{{x_j^b}\}, \forall {i \neq j}$. Finally, for each $x_i$, we can get a set of $2n-2$ negative samples.

For an input $x_i^a$ with representation $v_i^a$ under the view of $a$, it has one positive sample $x_i^b$ with representation $v_i^b$ under the view of $b$. It also has a negative sample set $\mathbf{V^-} = \{\boldsymbol{v}_1^-, \ldots, \boldsymbol{v}_{2n-2}^-\}$ with size $2n-2$, which consists of two types of negative sample subsets, e.g., intra-view negative sample set $\mathbf{V_1^-}$ with size $n-1$, where $\boldsymbol{v}_j^a \in \mathbf{V_1^-}, \forall {j \neq i}$, and the inter-view negative sample set $\mathbf{V_2^-}$  with size $n-1$, where $\boldsymbol{v}_j^b \in \mathbf{V_2^-}, \forall {j \neq i}$.  We define the similarity of a pair of samples as the dot product of their representations.
Then the loss function for a positive pair $(x_i^a, x_i^b)$ can be defined as:
\begin{equation}
 	\small
		\label{eq:cl} 
		l(x_i^a, x_i^b)\!=\!- {\rm ln} \frac{{\rm exp}(\boldsymbol v_i^a \cdot \boldsymbol{v}_i^b)}
		{
		{\rm exp}(\boldsymbol v_i^a \cdot \boldsymbol{v}_i^b)\! +\! \sum_{k=1}^{2n-2}{\rm exp}(\boldsymbol{v}_i^a \cdot \boldsymbol{v}_k^-)
		 }\,.\!
\end{equation}
We calculate the loss for the same pair twice with order switched, i.e., $(x_i^a, x_i^b)$ is changed to $(x_i^b, x_i^a)$ as the dot product with negative samples for $x_i^a$ and $x_i^b$ are different. Overall, the MVCL loss function is defined as follows:
\begin{equation}
	\label{eq:batch_cl} 
	\mathcal{L}_{\rm MVCL}\!=\!-\frac{1}{|\mathcal{N}|} \sum^{|\mathcal{N}|}_i \left[ l(x_i^a,x_i^b)\!+\!l(x_i^b,x_i^a) \right]\,,\!
\end{equation}
where $\mathcal{N}$ denotes the set of all program samples covering all different views.

\begin{figure}
	\centering
	\includegraphics[width=0.48\textwidth]{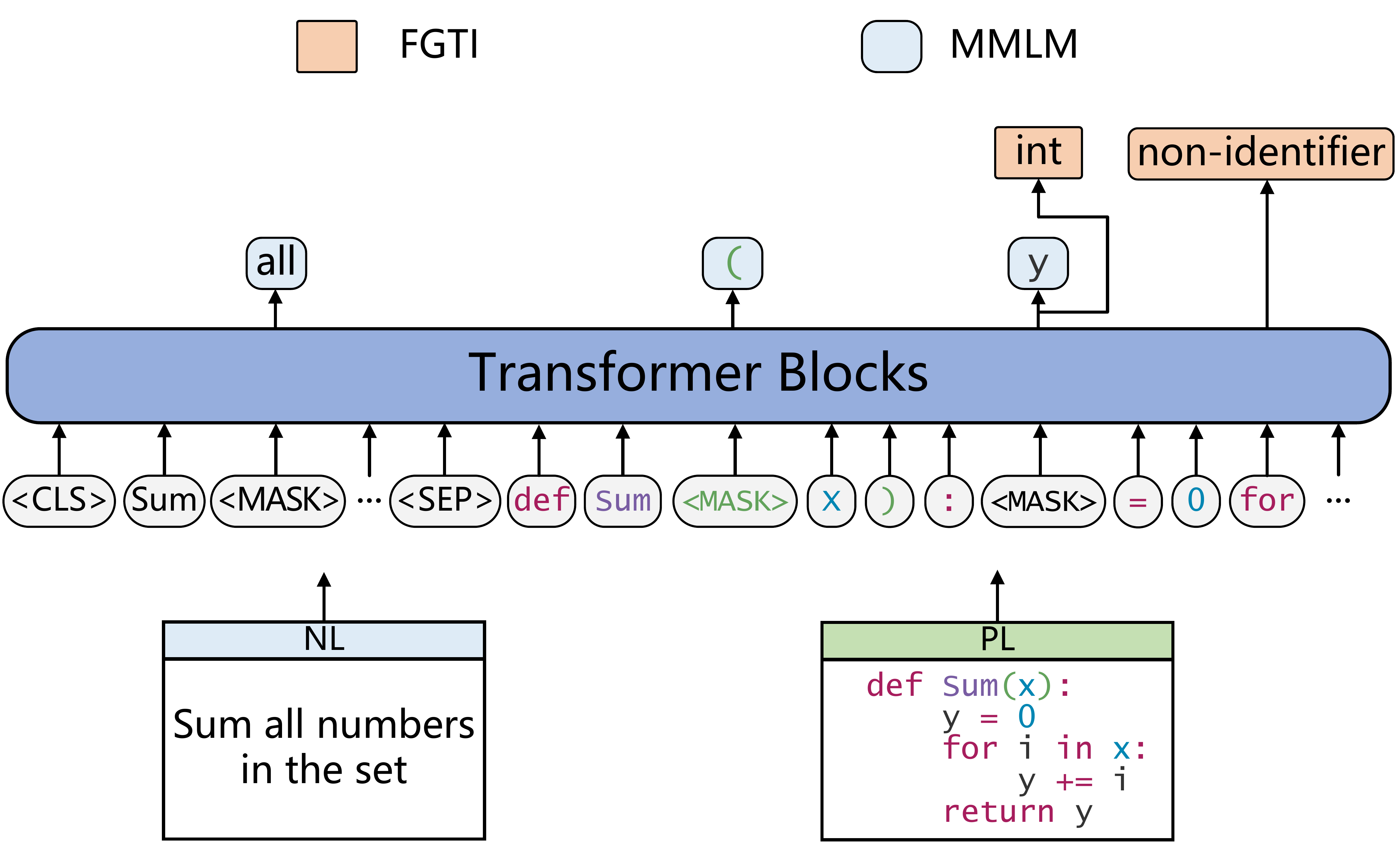}
	\caption{\label{fig:type}
		Pre-training with fine-grained type inference and multi-view masked language modeling.
	}
\end{figure}
\subsection{Pre-Training with Type Inference}
Figure \ref{fig:type} shows the other two pre-training tasks, including fine-grained type inference and multi-view masked language modeling.
\paragraph{Fine-Grained Type Inference.}
Several previous works~\cite{Wang2021SynCoBERTSM, Wang2021CodeT5IU} have proven the importance of symbolic properties in programming languages. Two concurrent works, SynCoBERT~\cite{Wang2021SynCoBERTSM} and CodeT5~\cite{Wang2021CodeT5IU} let the model divide the code token types into \texttt{identifier} or \texttt{non-identifier}. 
Inspired by the type checking in compilation process, we propose a fine-grained type inference (FGTI) objective to capture the fine-grained type information of variables~\cite{li2022, AnCFH11}. First, we parse all source codes into ASTs. Then, we traverse the AST and use the type checker to obtain fine-grained identifier types. We employ BPE tokenizer~\cite{Sennrich2016NeuralMT} to tokenize tokens and let sub-tokens inherit the type information of the token. Finally, we define the loss function as follows:
\begin{equation}
	\label{eq:fgtp}
	\mathcal{L}_{\rm FGTI} = -\frac{1}{|\mathcal{Z}|} \sum^{|\mathcal{Z}|}_i \sum^{|\mathcal{T}|}_j \ Y_{ij}\ {\rm log}\ P_{ij}\,,
\end{equation}
where $\mathcal{Z}$ denotes the set of all tokens that need to inference types, $\mathcal{T}$ represents the set of all types contained in the pre-training corpus, $Y_{ij}$ denotes the label of token $i$ in type $j$, and $P_{ij}$ denotes the predicted probability of token $i$ in type $j$.
\paragraph{Multi-View Masked Language Modeling.}
In addition to the multi-view contrastive learning objective and fine-grained type inference objective, we also extend the Masked Language Modeling (MLM) to the multi-view program corpus, named MMLM. Given a data point $x$, we randomly select 15\% of tokens in $x$ and replace them with a special token $\texttt{<MASK>}$, following the same settings in~\cite{devlin2018bert}. The MMLM objective aims to predict original tokens which are masked out. We calculate the MMLM loss as follows:
\begin{equation}
	\label{eq:mmlm}
	\mathcal{L}_{\rm MMLM} = -\frac{1}{|\mathcal{M}|} \sum^{|\mathcal{M}|}_i \sum^{|\mathcal{V}|}_j \ Y_{ij}\ {\rm log}\ P_{ij}\,,
\end{equation}
where $\mathcal{M}$ denotes the set of masked tokens, $\mathcal{V}$ represents the vocabulary, $Y_{ij}$ denotes the label of the masked token $i$ in class $j$, and $P_{ij}$ denotes the predicted probability of token $i$ in class $j$. 

\subsection{Overall Training Objective}
	The overall loss function in \ourapproach is the integration of several components we have defined before.
	\begin{equation}
    	\mathcal{L} = \mathcal{L}_{\rm MVCL} + \mathcal{L}_{\rm FGTI} + \mathcal{L}_{\rm MMLM} + \lambda \lVert \Theta\rVert^2\,,
	\end{equation}
	where $\Theta$ contains all trainable parameters of the model, and $\lambda$ is the coefficient of $L_2$ regularizer.

\section{Experimental Setup}
We conduct experiments to answer the following research questions: (1) \textit{How effective is \ourapproach compared with the state-of-the-art baselines?} (2) \textit{How do different components and different views affect our \ourapproach?}

\subsection{Pre-Training Dataset and Settings}
Different programming languages often require different program analyzers. Existing program analysis tools rarely support multiple programming languages and multi-view program transformations. For convenience, we choose Python for our experiments, as it is very popular and used in many projects. 
We pre-train \ourapproach on the Python corpus of CodeSearchNet dataset~\cite{Husain2019CodeSearchNetCE}, which consists of 0.5M bimodal Python functions with their corresponding natural-language comments, as well as 1.1M unimodal Python functions.

\ourapproach is built on the top of Transformer~\cite{vaswani2017attention}, and consists of a 12-layer encoder with 768 hidden sizes and 12 attention heads. The pre-training procedure is conducted on 8 NVIDIA V100 GPUs for 600K steps, with each mini-batch containing 128 sequences up to 512 tokens including special tokens. According to the length distribution of samples in the training corpus, we set the lengths of PL/AST/CFG/PT in unpaired data to 512, and set the lengths of NL and PL/AST/CFG/PT in paired data to 96 and 416 respectively. The learning rate of \ourapproach is set to $1e\textit{-}4$ with a linear warm up over the first 30K steps and a linear decay. \ourapproach is trained with a dropout rate of 0.1 on all layers and attention weights. We initialize the parameters of \ourapproach by GraphCodeBERT~\cite{Guo2021GraphCodeBERTPC} and utilize a BPE tokenizer~\cite{Sennrich2016NeuralMT}.

\begin{table}[!t]
	\small
	\centering
	\setlength{\tabcolsep}{0.8mm}{
	    \begin{tabular}{l c c c c}
		    \toprule
	        Tasks & Datasets &Train&Valid&Test \\
	        \midrule
	        \multirow{3}{2.5cm}{Natural Language Code Retrieval}&AdvTest&251K&9.6K&19.2K\\
	        &CosQA&19.6K&0.5K&0.5K\\
	        &CoNaLa&2.4K&-&0.5K\\
	        \midrule
	        Code-to-Code Retrieval&Python800&72K&4K&4K\\
	        Code Clone Detection&Python800&144K&8K&8K\\
	        \midrule
	        Code Defect Detection&GREAT&100K&5K&5K\\
			\bottomrule
		\end{tabular}
		\caption{Statistics of datasets for downstream tasks.} 
	    \label{table:ft_data}
	    }
\end{table}

\subsection{Evaluation Tasks, Datasets and Metrics}
We select several program comprehension tasks to evaluate \ourapproach, including natural language code retrieval, code similarity, and code defect detection. 
We pre-train \ourapproach on Python corpus, and choose several public Python datasets to evaluate it, as shown in Table~\ref{table:ft_data}.

\paragraph{Natural Language Code Retrieval.} This task aims to find the most relevant code snippet from a collection of candidates, given a natural language query. We choose three datasets to evaluate this task, including AdvTest~\cite{Lu2021CodeXGLUEAM}, CoNaLa~\cite{Yin2018LearningTM}, and CoSQA~\cite{Huang2021CoSQA20}. We adopt the Mean Reciprocal Rank (MRR) metric to evaluate the performance of code retrieval. In AdvTest dataset, we set the learning rate as $5e\text{-}5$, the batch size as 32, the maximum fine-tuning epoch as 20, the maximum length of both query and code sequence as 256. In CoNaLa and CoSQA datasets, we set the learning rate as $5e\text{-}5$, the batch size as 32, the maximum fine-tuning epoch as 30, the maximum length of query and code sequence as 128. In AdvTest and CoSQA datasets, we save the optimal checkpoint on the validation set, and test it on the testing set. In CoNaLa dataset, we report the best results on the testing set.

\paragraph{Code Similarity.} 
This task is always categorized into two groups: code-to-code retrieval and code clone detection. We conduct experiments on the Python800 dataset~\cite{Puri2021ProjectCA}, which is composed of 800 problems with each problem having 300 unique Python solution files. We remove those files not in UTF-8 encoding formats and randomly select 100 solutions for each problem. In code-to-code retrieval, the filtered dataset is split to 720/40/40 problems for training, validation, and testing. Given a program, this task aims to retrieve other programs that solve the same problem; we evaluate using Mean Average Precision (MAP).
Regarding the task of code clone detection, we treat it as binary classification and evaluate it using the Accuracy score, following~\cite{Puri2021ProjectCA}. 

To train these two tasks, we set the learning rate as $2e\text{-}5$, the batch size as 32, the epoch number as 20. In code-to-code retrieval, we set the maximum length of both query and code sequence as 256. In code clone detection, we set the maximum concatenation sequence length of the two code snippets to 512. We save the optimal checkpoint on the validation set, and test it on the testing set.

\paragraph{Code Defect Detection.} This task aims to identify whether a given piece of code snippet is vulnerable or not, which is usually treated as a binary classification task. We evaluate all models on the GREAT dataset~\cite{Hellendoorn2020GlobalRM}, which is originally built from the ETH Py150 dataset~\cite{Raychev2016ProbabilisticMF}. We evaluate the performance of code defect detection using the Accuracy score. We randomly select 100K samples for training, 5K samples for validation and 5K samples for testing, respectively. We set the learning rate as $5e\text{-}5$, the batch size as 32, the maximum fine-tuning epoch as 50, the maximum length of both query and code sequence as 256. We save the optimal checkpoint on the validation set, and test it on the testing set.

\subsection{Baselines}
We compare \ourapproach with various state-of-the-art models.
    \textbf{RoBERTa}~\cite{Liu2019RoBERTaAR} is a robustly optimized BERT~\cite{devlin2018bert}, which is originally pre-trained on a large-scale natural-language corpus. We fine-tune it on source code datasets of downstream tasks.
    \textbf{CodeBERT}~\cite{Feng2020CodeBERTAP} is pre-trained on NL-PL pairs using both masked language modeling~\cite{devlin2018bert} and replaced token detection~\cite{clark2020electra} objectives.
    \textbf{GraphCodeBERT}~\cite{Guo2021GraphCodeBERTPC} is a pre-trained language model of source code which incorporates the data flow information of source code.
    \textbf{PLBART}~\cite{ahmad2021unified} is based on the BART~\cite{lewis2020bart} architecture and pre-trained on Python and Java functions using denoising autoencoding.
    \textbf{CodeT5}~\cite{Wang2021CodeT5IU} is based on the T5~\cite{raffel2019exploring} architecture and employs denoising sequence-to-sequence pre-training on seven programming languages.
    \textbf{SynCoBERT}~\cite{Wang2021SynCoBERTSM} incorporates AST by edge prediction and uses contrastive learning to maximize the mutual information among programs, documents, and ASTs.

\begin{table}[!t]
	\small
	\centering
	\setlength{\tabcolsep}{1.3mm}{
	
		\begin{tabular}{lcccc}
			\toprule
			Models & AdvTest & CoNaLa & CoSQA & Average\\
			\midrule
			RoBERTa &18.3 &30.7 &57.6 & 35.5\\
			CodeBERT &27.2 &38.9 &64.2 & 43.4\\
			GraphCodeBERT &35.2 &47.3 &68.2 & 50.2\\ 
			PLBART &34.3 &45.5 &65.3 & 48.4\\
			CodeT5 &36.5 &47.7 &67.7 & 50.6\\
			SynCoBERT &38.1 &48.4 &69.6 & 52.0\\
			\ourapproach &\textbf{40.4} &\textbf{50.6} &\textbf{72.1} & \textbf{54.4} \\
			\bottomrule
    	\end{tabular}
    	\caption{Results on the natural language code retrieval task evaluating with MRR, using the AdvTest, CoNaLa, and CoSQA datasets.} 
		\label{table:code_retrieval}
		}
\end{table}
\section{Results and Analysis}
\subsection{Performance on Downstream Tasks (RQ1)}
\paragraph{Natural Language Code Retrieval.}
Table~\ref{table:code_retrieval} shows the results of natural language code retrieval on three datasets. We can observe that \ourapproach outperforms all baseline models on all datasets. Specifically, it outperforms CodeT5 by 3.8 points on average. Compared to the previous state-of-the-art SynCoBERT, \ourapproach also performs better with an average improvement of 2.4 points.
This significant performance improvement indicates that the code representation learned by \ourapproach preserves more code semantics.
We attribute this improvement to our introduced multi-view contrastive pre-training strategy. 

\begin{table}[!t]
	\small
	\centering
	\setlength{\tabcolsep}{2mm}{
		\begin{tabular}{lcc}
			\toprule
			Models  &MAP@R &Accuracy\\
			\midrule
			RoBERTa &82.9&94.4\\
			CodeBERT &86.1&95.2\\
			GraphCodeBERT &88.8&95.9\\ 
			PLBART &86.7&95.5\\
			CodeT5 &88.1&95.7\\
			SynCoBERT &89.2&96.1\\
			\ourapproach &\textbf{91.5}&\textbf{97.4}\\
			\bottomrule
	    \end{tabular}
	    \caption{Results on the code-to-code retrieval and code clone detection tasks evaluating with MAP and Accuracy score, using the Python800 dataset.} 
		\label{table:code_sim}
	    }
\end{table}
\paragraph{Code Similarity.}
Table~\ref{table:code_sim} presents the results for code similarity calculation, including code-to-code retrieval and code clone detection. We can see that \ourapproach significantly outperforms all baseline models on these two tasks. In the task of code-to-code retrieval, \ourapproach outperforms CodeT5 and SynCoBERT by 3.4 points and 2.3 points, respectively. In the task of code clone detection, \ourapproach achieves 1.5 and 1.3 points higher compared to GraphCodeBERT and SynCoBERT, respectively. These results show that \ourapproach can better identify those programs with the same semantics and distinguish those programs with different semantics.

\paragraph{Code Defect Detection.}
\begin{table}[!t]
	\small
	\centering
		
		\begin{tabular}{lc}
			\toprule
			Models & Accuracy\\
			\midrule
			RoBERTa &81.9\\
			CodeBERT &85.5\\
			GraphCodeBERT &87.5\\ 
			PLBART &86.8\\
			CodeT5 &87.4\\
			SynCoBERT &88.2\\
			\ourapproach &\textbf{89.3}\\
			\bottomrule
	    \end{tabular}
	    \caption{Results on the code defect detection task evaluating with Accuracy score, using the GREAT dataset.} 
		\label{table:defect_detection}
\end{table}
Table~\ref{table:defect_detection} shows the experimental results of code defect detection. \ourapproach consistently outperforms all models. Specifically, it outperforms GraphCodeBERT and SynCoBERT by 1.8 and 1.1 points, respectively. These results indicate that \ourapproach can effectively preserve the semantics of programs, which is beneficial for code defect detection.

\subsection{Ablation Study (RQ2)}
We empirically study several simplified variants of \ourapproach to understand the contributions of each component, including the Multi-View Contrastive Learning (MVCL), Fine-Grained Type Inference (FGTI), Abstract Syntax Tree (AST), Program Transformation (PT), and Control Flow Graph (CFG). 
Taking the natural language code retrieval task as an example, Table~\ref{table:ablation} shows the experimental results of each variant on that task.
The setting of w/o (MVCL, FGTI) indicates that these pre-training objectives are removed from \ourapproach respectively. The setting of w/o (AST, PT, CFG) indicates that different views of programs are removed from \ourapproach respectively. From Table~\ref{table:ablation}, several meaningful observations can be drawn. (1) Both MVCL and FGTI effectively increase the performance, which confirms that the two proposed pre-training objectives can indeed improve the ability of the model for program comprehension. (2) Exploiting different views of programs can bring performance improvements to the model as arbitrarily discarding any view of programs degrades the performance. Additionally, the introduction of CFG brings more performance improvements, indicating the importance of execution information for program understanding.

\begin{table}[!t]
	\small
	\centering
	\setlength{\tabcolsep}{1.8mm}{
	
		\begin{tabular}{lcccc}
			\toprule
			Models & AdvTest & CoNaLa & CoSQA & Average\\
			\hline
			\ourapproach &\textbf{40.4} &\textbf{50.6} &\textbf{72.1} & \textbf{54.4} \\
			\ \ \ w/o MVCL &36.2 &47.7 &69.2 &51.0\\
			\ \ \ w/o FGTI &38.0 &48.9 &70.8 &52.6\\
			\hdashline
			\ \ \ w/o AST &39.1 &48.5 &71.3 &53.0\\
			\ \ \ w/o PT  &38.2 &48.6 &70.8 &52.5\\
			\ \ \ w/o CFG &37.8 &47.9 &70.5 &52.1\\
			\bottomrule
    	\end{tabular}
    	\caption{Ablation study on the task of natural language code retrieval, evaluated using MRR.} 
		\label{table:ablation}
		}
\end{table}

\section{Related Work}
\paragraph{Pre-Trained Models for Source Code.}
Benefiting from the strong power of pre-trained models in natural language processing~\cite{Liu2019RoBERTaAR, devlin2018bert, Wang21ICSOC, Wang20ASE, Wang20SS}, several recent works attempt to use the pre-training techniques on programs~\cite{svyatkovskiy2020intellicode}. \citet{kanade2020learning} proposed CuBERT which follows the architecture of BERT~\cite{devlin2018bert}, and is pre-trained with a masked language modeling objective on a large-scale Python corpus. 
\citet{Feng2020CodeBERTAP} proposed CodeBERT, which is pre-trained on NL-PL pairs in six programming languages, introducing the replaced token detection objective~\cite{clark2020electra}. Furthermore, \citet{Guo2021GraphCodeBERTPC} proposed GraphCodeBERT, which incorporates the data flow of programs into the model pre-training process. 
\citet{Wang2021SynCoBERTSM} proposed SynCoBERT, which incorporates ASTs via edge prediction to enhance the structural information of programs. They also used contrastive learning to maximize the mutual information among programs, documents, and ASTs.
\citet{Lu2021CodeXGLUEAM} proposed CodeGPT for code completion, which is pre-trained using a unidirectional language modeling objective. \citet{ahmad2021unified} proposed PLBART based on BART~\cite{lewis2020bart}, which is pre-trained on a large-scale corpus of Java and Python programs paired with their corresponding comments via denoising autoencoding. \citet{Wang2021CodeT5IU} proposed CodeT5 following the architecture of T5~\cite{raffel2019exploring}. It employs denoising sequence-to-sequence pre-training on seven programming languages.
Recently,~\citet{wan2022they} conducted a thorough structural analysis aiming to provide an interpretation of pre-trained language models for source code (e.g., CodeBERT and GraphCodeBERT).

\paragraph{Program Analysis for Code Intelligence.}
In addition to the lexical information of programs, many recent works attempt to leverage program analysis techniques to capture the structural and syntactic representations of programs~\cite{Cummins2020ProGraMLGD}. \citet{Kim2021CodePB} designed several strategies to feed the ASTs of programs into Transformer~\cite{vaswani2017attention}.
\citet{Li2019GraphMN} proposed a graph matching network, which utilizes the CFG of the program to deal with the challenge of binary function similarity search. 
\citet{LingWWPMXLWJ21} proposed a deep graph matching and searching model based on graph neural networks~\cite{KipfW17, Wangbundle21, Wangicwspoi, YuZLLXW22, ZHAO2022111324} for code retrieval. They represented both natural language queries and code snippets based on the unified graph-structured data.
\citet{Iyer2020SoftwareLC} presented the program-derived semantic graph to capture the semantics of programs at multiple levels of abstraction.
\citet{BenNun2018NeuralCC} presented inst2vec, which locally embeds individual statement in LLVM intermediate representations by processing a contextual flow graph with a context prediction objective~\cite{Mikolov2013DistributedRO}.

\paragraph{Contrastive Learning on Programs.}
Recently, several attempts have been made to leverage contrastive learning for better code semantics. ContraCode~\cite{jain2021contrastive} and Corder~\cite{bui2021self} first utilized semantic-preserving program transformations such as identifier renaming, dead code insertion, to build positive instances. Then a contrastive learning objective is designed to maximize the mutual information among the positive and negative instances.
\citet{Ding2021ContrastiveLF} presented a self-supervised pre-training technique called BOOST based on contrastive learning. They inject real-world bugs to build hard negative pairs. 
In \ourapproach, we construct the positive pairs throughout the compilation process of programs, including lexical analysis, syntax analysis, semantic analysis, and static analysis. It is the first pre-trained model that integrates multi-views of programs for program comprehension.

\section{Conclusion} 
In this paper, we have proposed \ourapproach, a novel approach to represent the source code with multi-view contrastive pre-training learning.
We extract multiple code views with compiler tools and learn the complement among them under a contrastive learning framework. We also propose a fine-grained type inference task in the pre-training process.
Comprehensive experiments on three downstream tasks over five datasets verify the effectiveness of \ourapproach when compared with several state-of-the-art baselines.

\section*{Acknowledgements}
We would like to thank Gerasimos Lampouras and Ignacio Iacobacci from Huawei London Research Institute for their constructive comments on this paper.
Jin Liu is supported by National Natural Science Foundation of China under Grant No. 61972290. 
Yao Wan is partially supported by National Natural Science Foundation of China under Grant No. 62102157. Hao Wu is supported by National Natural Science Foundation of China under Grant No. 61962061, and partially supported by Yunnan Provincial Foundation for Leaders of Disciplines in Science and Technology (202005AC160005).
\bibliography{ref}

\begin{thebibliography}{63}
\expandafter\ifx\csname natexlab\endcsname\relax\def\natexlab#1{#1}\fi

\bibitem[{Ahmad et~al.(2021)Ahmad, Chakraborty, Ray, and
  Chang}]{ahmad2021unified}
Wasi~Uddin Ahmad, Saikat Chakraborty, Baishakhi Ray, and Kai{-}Wei Chang. 2021.
\newblock \href {https://doi.org/10.18653/v1/2021.naacl-main.211} {Unified
  pre-training for program understanding and generation}.
\newblock In \emph{Proceedings of the 2021 Conference of the North American
  Chapter of the Association for Computational Linguistics: Human Language
  Technologies, {NAACL-HLT} 2021, Online, June 6-11, 2021}, pages 2655--2668.
  Association for Computational Linguistics.

\bibitem[{Allamanis et~al.(2018)Allamanis, Barr, Devanbu, and
  Sutton}]{allamanis2018survey}
Miltiadis Allamanis, Earl~T. Barr, Premkumar~T. Devanbu, and Charles Sutton.
  2018.
\newblock \href {https://doi.org/10.1145/3212695} {A survey of machine learning
  for big code and naturalness}.
\newblock \emph{{ACM} Comput. Surv.}, 51(4):81:1--81:37.

\bibitem[{Alon et~al.(2019)Alon, Zilberstein, Levy, and Yahav}]{AlonZLY19}
Uri Alon, Meital Zilberstein, Omer Levy, and Eran Yahav. 2019.
\newblock \href {https://doi.org/10.1145/3290353} {code2vec: learning
  distributed representations of code}.
\newblock \emph{Proc. {ACM} Program. Lang.}, 3({POPL}):40:1--40:29.

\bibitem[{An et~al.(2011)An, Chaudhuri, Foster, and Hicks}]{AnCFH11}
Jong{-}hoon~(David) An, Avik Chaudhuri, Jeffrey~S. Foster, and Michael Hicks.
  2011.
\newblock \href {https://doi.org/10.1145/1926385.1926437} {Dynamic inference of
  static types for ruby}.
\newblock In \emph{Proceedings of the 38th {ACM} {SIGPLAN-SIGACT} Symposium on
  Principles of Programming Languages, {POPL} 2011, Austin, TX, USA, January
  26-28, 2011}, pages 459--472. {ACM}.

\bibitem[{Ben{-}Nun et~al.(2018)Ben{-}Nun, Jakobovits, and
  Hoefler}]{BenNun2018NeuralCC}
Tal Ben{-}Nun, Alice~Shoshana Jakobovits, and Torsten Hoefler. 2018.
\newblock \href
  {https://proceedings.neurips.cc/paper/2018/hash/17c3433fecc21b57000debdf7ad5c930-Abstract.html}
  {Neural code comprehension: {A} learnable representation of code semantics}.
\newblock In \emph{Advances in Neural Information Processing Systems 31: Annual
  Conference on Neural Information Processing Systems 2018, NeurIPS 2018,
  December 3-8, 2018, Montr{\'{e}}al, Canada}, pages 3589--3601.

\bibitem[{Bui et~al.(2021{\natexlab{a}})Bui, Yu, and
  Jiang}]{Bui2021InferCodeSL}
Nghi D.~Q. Bui, Yijun Yu, and Lingxiao Jiang. 2021{\natexlab{a}}.
\newblock \href {https://doi.org/10.1109/ICSE43902.2021.00109} {Infercode:
  Self-supervised learning of code representations by predicting subtrees}.
\newblock In \emph{43rd {IEEE/ACM} International Conference on Software
  Engineering, {ICSE} 2021, Madrid, Spain, 22-30 May 2021}, pages 1186--1197.
  {IEEE}.

\bibitem[{Bui et~al.(2021{\natexlab{b}})Bui, Yu, and Jiang}]{bui2021self}
Nghi D.~Q. Bui, Yijun Yu, and Lingxiao Jiang. 2021{\natexlab{b}}.
\newblock \href {https://doi.org/10.1145/3404835.3462840} {Self-supervised
  contrastive learning for code retrieval and summarization via
  semantic-preserving transformations}.
\newblock In \emph{{SIGIR} '21: The 44th International {ACM} {SIGIR} Conference
  on Research and Development in Information Retrieval, Virtual Event, Canada,
  July 11-15, 2021}, pages 511--521. {ACM}.

\bibitem[{Chen et~al.(2020)Chen, Kornblith, Norouzi, and Hinton}]{Chen2020ASF}
Ting Chen, Simon Kornblith, Mohammad Norouzi, and Geoffrey~E. Hinton. 2020.
\newblock \href {http://proceedings.mlr.press/v119/chen20j.html} {A simple
  framework for contrastive learning of visual representations}.
\newblock In \emph{Proceedings of the 37th International Conference on Machine
  Learning, {ICML} 2020, 13-18 July 2020, Virtual Event}, volume 119 of
  \emph{Proceedings of Machine Learning Research}, pages 1597--1607. {PMLR}.

\bibitem[{Clark et~al.(2020)Clark, Luong, Le, and Manning}]{clark2020electra}
Kevin Clark, Minh{-}Thang Luong, Quoc~V. Le, and Christopher~D. Manning. 2020.
\newblock \href {https://openreview.net/forum?id=r1xMH1BtvB} {{ELECTRA:}
  pre-training text encoders as discriminators rather than generators}.
\newblock In \emph{8th International Conference on Learning Representations,
  {ICLR} 2020, Addis Ababa, Ethiopia, April 26-30, 2020}. OpenReview.net.

\bibitem[{Cummins et~al.(2020)Cummins, Fisches, Ben{-}Nun, Hoefler, and
  Leather}]{Cummins2020ProGraMLGD}
Chris Cummins, Zacharias~V. Fisches, Tal Ben{-}Nun, Torsten Hoefler, and Hugh
  Leather. 2020.
\newblock \href {http://arxiv.org/abs/2003.10536} {Programl: Graph-based deep
  learning for program optimization and analysis}.
\newblock \emph{CoRR}, abs/2003.10536.

\bibitem[{Devlin et~al.(2019)Devlin, Chang, Lee, and
  Toutanova}]{devlin2018bert}
Jacob Devlin, Ming{-}Wei Chang, Kenton Lee, and Kristina Toutanova. 2019.
\newblock \href {https://doi.org/10.18653/v1/n19-1423} {{BERT:} pre-training of
  deep bidirectional transformers for language understanding}.
\newblock In \emph{Proceedings of the 2019 Conference of the North American
  Chapter of the Association for Computational Linguistics: Human Language
  Technologies, {NAACL-HLT} 2019, Minneapolis, MN, USA, June 2-7, 2019, Volume
  1 (Long and Short Papers)}, pages 4171--4186. Association for Computational
  Linguistics.

\bibitem[{Ding et~al.(2021)Ding, Buratti, Pujar, Morari, Ray, and
  Chakraborty}]{Ding2021ContrastiveLF}
Yangruibo Ding, Luca Buratti, Saurabh Pujar, Alessandro Morari, Baishakhi Ray,
  and Saikat Chakraborty. 2021.
\newblock \href {http://arxiv.org/abs/2110.03868} {Contrastive learning for
  source code with structural and functional properties}.
\newblock \emph{CoRR}, abs/2110.03868.

\bibitem[{Feng et~al.(2020)Feng, Guo, Tang, Duan, Feng, Gong, Shou, Qin, Liu,
  Jiang, and Zhou}]{Feng2020CodeBERTAP}
Zhangyin Feng, Daya Guo, Duyu Tang, Nan Duan, Xiaocheng Feng, Ming Gong, Linjun
  Shou, Bing Qin, Ting Liu, Daxin Jiang, and Ming Zhou. 2020.
\newblock \href {https://doi.org/10.18653/v1/2020.findings-emnlp.139}
  {Codebert: {A} pre-trained model for programming and natural languages}.
\newblock In \emph{Findings of the Association for Computational Linguistics:
  {EMNLP} 2020, Online Event, 16-20 November 2020}, volume {EMNLP} 2020 of
  \emph{Findings of {ACL}}, pages 1536--1547. Association for Computational
  Linguistics.

\bibitem[{Guo et~al.(2021)Guo, Ren, Lu, Feng, Tang, Liu, Zhou, Duan,
  Svyatkovskiy, Fu, Tufano, Deng, Clement, Drain, Sundaresan, Yin, Jiang, and
  Zhou}]{Guo2021GraphCodeBERTPC}
Daya Guo, Shuo Ren, Shuai Lu, Zhangyin Feng, Duyu Tang, Shujie Liu, Long Zhou,
  Nan Duan, Alexey Svyatkovskiy, Shengyu Fu, Michele Tufano, Shao~Kun Deng,
  Colin~B. Clement, Dawn Drain, Neel Sundaresan, Jian Yin, Daxin Jiang, and
  Ming Zhou. 2021.
\newblock \href {https://openreview.net/forum?id=jLoC4ez43PZ} {Graphcodebert:
  Pre-training code representations with data flow}.
\newblock In \emph{9th International Conference on Learning Representations,
  {ICLR} 2021, Virtual Event, Austria, May 3-7, 2021}. OpenReview.net.

\bibitem[{Hellendoorn et~al.(2020)Hellendoorn, Sutton, Singh, Maniatis, and
  Bieber}]{Hellendoorn2020GlobalRM}
Vincent~J. Hellendoorn, Charles Sutton, Rishabh Singh, Petros Maniatis, and
  David Bieber. 2020.
\newblock \href {https://openreview.net/forum?id=B1lnbRNtwr} {Global relational
  models of source code}.
\newblock In \emph{8th International Conference on Learning Representations,
  {ICLR} 2020, Addis Ababa, Ethiopia, April 26-30, 2020}. OpenReview.net.

\bibitem[{Huang et~al.(2021)Huang, Tang, Shou, Gong, Xu, Jiang, Zhou, and
  Duan}]{Huang2021CoSQA20}
Junjie Huang, Duyu Tang, Linjun Shou, Ming Gong, Ke~Xu, Daxin Jiang, Ming Zhou,
  and Nan Duan. 2021.
\newblock \href {https://doi.org/10.18653/v1/2021.acl-long.442} {Cosqa: 20,
  000+ web queries for code search and question answering}.
\newblock In \emph{Proceedings of the 59th Annual Meeting of the Association
  for Computational Linguistics and the 11th International Joint Conference on
  Natural Language Processing, {ACL/IJCNLP} 2021, (Volume 1: Long Papers),
  Virtual Event, August 1-6, 2021}, pages 5690--5700. Association for
  Computational Linguistics.

\bibitem[{Husain et~al.(2019)Husain, Wu, Gazit, Allamanis, and
  Brockschmidt}]{Husain2019CodeSearchNetCE}
Hamel Husain, Ho{-}Hsiang Wu, Tiferet Gazit, Miltiadis Allamanis, and Marc
  Brockschmidt. 2019.
\newblock \href {http://arxiv.org/abs/1909.09436} {Codesearchnet challenge:
  Evaluating the state of semantic code search}.
\newblock \emph{CoRR}, abs/1909.09436.

\bibitem[{Iyer et~al.(2020)Iyer, Sun, Wang, and
  Gottschlich}]{Iyer2020SoftwareLC}
Roshni~G. Iyer, Yizhou Sun, Wei Wang, and Justin Gottschlich. 2020.
\newblock \href {http://arxiv.org/abs/2004.00768} {Software language
  comprehension using a program-derived semantic graph}.
\newblock \emph{CoRR}, abs/2004.00768.

\bibitem[{Iyer et~al.(2016)Iyer, Konstas, Cheung, and
  Zettlemoyer}]{iyer2016summarizing}
Srinivasan Iyer, Ioannis Konstas, Alvin Cheung, and Luke Zettlemoyer. 2016.
\newblock \href {https://doi.org/10.18653/v1/p16-1195} {Summarizing source code
  using a neural attention model}.
\newblock In \emph{Proceedings of the 54th Annual Meeting of the Association
  for Computational Linguistics, {ACL} 2016, August 7-12, 2016, Berlin,
  Germany, Volume 1: Long Papers}. The Association for Computer Linguistics.

\bibitem[{Jain et~al.(2021)Jain, Jain, Zhang, Abbeel, Gonzalez, and
  Stoica}]{jain2021contrastive}
Paras Jain, Ajay Jain, Tianjun Zhang, Pieter Abbeel, Joseph Gonzalez, and Ion
  Stoica. 2021.
\newblock \href {https://doi.org/10.18653/v1/2021.emnlp-main.482} {Contrastive
  code representation learning}.
\newblock In \emph{Proceedings of the 2021 Conference on Empirical Methods in
  Natural Language Processing, {EMNLP} 2021, Virtual Event / Punta Cana,
  Dominican Republic, 7-11 November, 2021}, pages 5954--5971. Association for
  Computational Linguistics.

\bibitem[{Javed et~al.(2004)Javed, Bryant, Crepinsek, Mernik, and
  Sprague}]{Javed2004ContextfreeGI}
Faizan Javed, Barrett~R. Bryant, Matej Crepinsek, Marjan Mernik, and Alan~P.
  Sprague. 2004.
\newblock \href {https://doi.org/10.1145/986537.986635} {Context-free grammar
  induction using genetic programming}.
\newblock In \emph{Proceedings of the 42nd Annual Southeast Regional
  Conference, 2004, Huntsville, Alabama, USA, April 2-3, 2004}, pages 404--405.
  {ACM}.

\bibitem[{Jiang et~al.(2021)Jiang, Zheng, Lyu, Li, and
  Lyu}]{Jiang2021TreeBERTAT}
Xue Jiang, Zhuoran Zheng, Chen Lyu, Liang Li, and Lei Lyu. 2021.
\newblock \href {https://proceedings.mlr.press/v161/jiang21a.html} {Treebert:
  {A} tree-based pre-trained model for programming language}.
\newblock In \emph{Proceedings of the Thirty-Seventh Conference on Uncertainty
  in Artificial Intelligence, {UAI} 2021, Virtual Event, 27-30 July 2021},
  volume 161 of \emph{Proceedings of Machine Learning Research}, pages 54--63.
  {AUAI} Press.

\bibitem[{Kanade et~al.(2020)Kanade, Maniatis, Balakrishnan, and
  Shi}]{kanade2020learning}
Aditya Kanade, Petros Maniatis, Gogul Balakrishnan, and Kensen Shi. 2020.
\newblock \href {http://proceedings.mlr.press/v119/kanade20a.html} {Learning
  and evaluating contextual embedding of source code}.
\newblock In \emph{Proceedings of the 37th International Conference on Machine
  Learning, {ICML} 2020, 13-18 July 2020, Virtual Event}, volume 119 of
  \emph{Proceedings of Machine Learning Research}, pages 5110--5121. {PMLR}.

\bibitem[{Kim et~al.(2021)Kim, Zhao, Tian, and Chandra}]{Kim2021CodePB}
Seohyun Kim, Jinman Zhao, Yuchi Tian, and Satish Chandra. 2021.
\newblock \href {https://doi.org/10.1109/ICSE43902.2021.00026} {Code prediction
  by feeding trees to transformers}.
\newblock In \emph{43rd {IEEE/ACM} International Conference on Software
  Engineering, {ICSE} 2021, Madrid, Spain, 22-30 May 2021}, pages 150--162.
  {IEEE}.

\bibitem[{Kipf and Welling(2017)}]{KipfW17}
Thomas~N. Kipf and Max Welling. 2017.
\newblock \href {https://openreview.net/forum?id=SJU4ayYgl} {Semi-supervised
  classification with graph convolutional networks}.
\newblock In \emph{5th International Conference on Learning Representations,
  {ICLR} 2017, Toulon, France, April 24-26, 2017, Conference Track
  Proceedings}. OpenReview.net.

\bibitem[{Lewis et~al.(2020)Lewis, Liu, Goyal, Ghazvininejad, Mohamed, Levy,
  Stoyanov, and Zettlemoyer}]{lewis2020bart}
Mike Lewis, Yinhan Liu, Naman Goyal, Marjan Ghazvininejad, Abdelrahman Mohamed,
  Omer Levy, Veselin Stoyanov, and Luke Zettlemoyer. 2020.
\newblock \href {https://doi.org/10.18653/v1/2020.acl-main.703} {{BART:}
  denoising sequence-to-sequence pre-training for natural language generation,
  translation, and comprehension}.
\newblock In \emph{Proceedings of the 58th Annual Meeting of the Association
  for Computational Linguistics, {ACL} 2020, Online, July 5-10, 2020}, pages
  7871--7880. Association for Computational Linguistics.

\bibitem[{Li et~al.(2022)Li, Wang, and Quan}]{li2022}
Li~Li, Jiawei Wang, and Haowei Quan. 2022.
\newblock \href {http://arxiv.org/abs/2202.11840} {Scalpel: The python static
  analysis framework}.
\newblock \emph{CoRR}, abs/2202.11840.

\bibitem[{Li et~al.(2019)Li, Gu, Dullien, Vinyals, and Kohli}]{Li2019GraphMN}
Yujia Li, Chenjie Gu, Thomas Dullien, Oriol Vinyals, and Pushmeet Kohli. 2019.
\newblock \href {http://proceedings.mlr.press/v97/li19d.html} {Graph matching
  networks for learning the similarity of graph structured objects}.
\newblock In \emph{Proceedings of the 36th International Conference on Machine
  Learning, {ICML} 2019, 9-15 June 2019, Long Beach, California, {USA}},
  volume~97 of \emph{Proceedings of Machine Learning Research}, pages
  3835--3845. {PMLR}.

\bibitem[{Ling et~al.(2021)Ling, Wu, Wang, Pan, Ma, Xu, Liu, Wu, and
  Ji}]{LingWWPMXLWJ21}
Xiang Ling, Lingfei Wu, Saizhuo Wang, Gaoning Pan, Tengfei Ma, Fangli Xu,
  Alex~X. Liu, Chunming Wu, and Shouling Ji. 2021.
\newblock \href {https://doi.org/10.1145/3447571} {Deep graph matching and
  searching for semantic code retrieval}.
\newblock \emph{{ACM} Trans. Knowl. Discov. Data}, 15(5):88:1--88:21.

\bibitem[{Liu et~al.(2019)Liu, Ott, Goyal, Du, Joshi, Chen, Levy, Lewis,
  Zettlemoyer, and Stoyanov}]{Liu2019RoBERTaAR}
Yinhan Liu, Myle Ott, Naman Goyal, Jingfei Du, Mandar Joshi, Danqi Chen, Omer
  Levy, Mike Lewis, Luke Zettlemoyer, and Veselin Stoyanov. 2019.
\newblock \href {http://arxiv.org/abs/1907.11692} {Roberta: {A} robustly
  optimized {BERT} pretraining approach}.
\newblock \emph{CoRR}, abs/1907.11692.

\bibitem[{Lu et~al.(2021)Lu, Guo, Ren, Huang, Svyatkovskiy, Blanco, Clement,
  Drain, Jiang, Tang, Li, Zhou, Shou, Zhou, Tufano, Gong, Zhou, Duan,
  Sundaresan, Deng, Fu, and Liu}]{Lu2021CodeXGLUEAM}
Shuai Lu, Daya Guo, Shuo Ren, Junjie Huang, Alexey Svyatkovskiy, Ambrosio
  Blanco, Colin~B. Clement, Dawn Drain, Daxin Jiang, Duyu Tang, Ge~Li, Lidong
  Zhou, Linjun Shou, Long Zhou, Michele Tufano, Ming Gong, Ming Zhou, Nan Duan,
  Neel Sundaresan, Shao~Kun Deng, Shengyu Fu, and Shujie Liu. 2021.
\newblock \href {http://arxiv.org/abs/2102.04664} {Codexglue: {A} machine
  learning benchmark dataset for code understanding and generation}.
\newblock \emph{CoRR}, abs/2102.04664.

\bibitem[{Mendis et~al.(2019)Mendis, Renda, Amarasinghe, and
  Carbin}]{MendisRAC19}
Charith Mendis, Alex Renda, Saman~P. Amarasinghe, and Michael Carbin. 2019.
\newblock \href {http://proceedings.mlr.press/v97/mendis19a.html} {Ithemal:
  Accurate, portable and fast basic block throughput estimation using deep
  neural networks}.
\newblock In \emph{Proceedings of the 36th International Conference on Machine
  Learning, {ICML} 2019, 9-15 June 2019, Long Beach, California, {USA}},
  volume~97 of \emph{Proceedings of Machine Learning Research}, pages
  4505--4515. {PMLR}.

\bibitem[{Mikolov et~al.(2013)Mikolov, Sutskever, Chen, Corrado, and
  Dean}]{Mikolov2013DistributedRO}
Tom{\'{a}}s Mikolov, Ilya Sutskever, Kai Chen, Gregory~S. Corrado, and Jeffrey
  Dean. 2013.
\newblock \href
  {https://proceedings.neurips.cc/paper/2013/hash/9aa42b31882ec039965f3c4923ce901b-Abstract.html}
  {Distributed representations of words and phrases and their
  compositionality}.
\newblock In \emph{Advances in Neural Information Processing Systems 26: 27th
  Annual Conference on Neural Information Processing Systems 2013. Proceedings
  of a meeting held December 5-8, 2013, Lake Tahoe, Nevada, United States},
  pages 3111--3119.

\bibitem[{Omri and Sinz(2020)}]{OmriS20}
Safa Omri and Carsten Sinz. 2020.
\newblock \href {https://doi.org/10.1145/3387940.3391463} {Deep learning for
  software defect prediction: {A} survey}.
\newblock In \emph{{ICSE} '20: 42nd International Conference on Software
  Engineering, Workshops, Seoul, Republic of Korea, 27 June - 19 July, 2020},
  pages 209--214. {ACM}.

\bibitem[{Paakki(1995)}]{Paakki1995AttributeGP}
Jukka Paakki. 1995.
\newblock \href {https://doi.org/10.1145/210376.197409} {Attribute grammar
  paradigms - {A} high-level methodology in language implementation}.
\newblock \emph{{ACM} Comput. Surv.}, 27(2):196--255.

\bibitem[{Phan et~al.(2021)Phan, Tran, Le, Nguyen, Anibal, Peltekian, and
  Ye}]{Phan2021CoTexTML}
Long~N. Phan, Hieu Tran, Daniel Le, Hieu Nguyen, James~T. Anibal, Alec
  Peltekian, and Yanfang Ye. 2021.
\newblock \href {http://arxiv.org/abs/2105.08645} {Cotext: Multi-task learning
  with code-text transformer}.
\newblock \emph{CoRR}, abs/2105.08645.

\bibitem[{Puri et~al.(2021)Puri, Kung, Janssen, Zhang, Domeniconi, Zolotov,
  Dolby, Chen, Choudhury, Decker, Thost, Buratti, Pujar, and
  Finkler}]{Puri2021ProjectCA}
Ruchir Puri, David~S. Kung, Geert Janssen, Wei Zhang, Giacomo Domeniconi,
  Vladimir Zolotov, Julian Dolby, Jie Chen, Mihir~R. Choudhury, Lindsey Decker,
  Veronika Thost, Luca Buratti, Saurabh Pujar, and Ulrich Finkler. 2021.
\newblock \href {http://arxiv.org/abs/2105.12655} {Project codenet: {A}
  large-scale {AI} for code dataset for learning a diversity of coding tasks}.
\newblock \emph{CoRR}, abs/2105.12655.

\bibitem[{Rabin et~al.(2020)Rabin, Bui, Yu, Jiang, and Alipour}]{Rabin2020OnTG}
Md. Rafiqul~Islam Rabin, Nghi D.~Q. Bui, Yijun Yu, Lingxiao Jiang, and
  Mohammad~Amin Alipour. 2020.
\newblock \href {http://arxiv.org/abs/2008.01566} {On the generalizability of
  neural program analyzers with respect to semantic-preserving program
  transformations}.
\newblock \emph{CoRR}, abs/2008.01566.

\bibitem[{Raffel et~al.(2020)Raffel, Shazeer, Roberts, Lee, Narang, Matena,
  Zhou, Li, and Liu}]{raffel2019exploring}
Colin Raffel, Noam Shazeer, Adam Roberts, Katherine Lee, Sharan Narang, Michael
  Matena, Yanqi Zhou, Wei Li, and Peter~J. Liu. 2020.
\newblock \href {http://jmlr.org/papers/v21/20-074.html} {Exploring the limits
  of transfer learning with a unified text-to-text transformer}.
\newblock \emph{J. Mach. Learn. Res.}, 21:140:1--140:67.

\bibitem[{Raychev et~al.(2016)Raychev, Bielik, and
  Vechev}]{Raychev2016ProbabilisticMF}
Veselin Raychev, Pavol Bielik, and Martin~T. Vechev. 2016.
\newblock \href {https://doi.org/10.1145/2983990.2984041} {Probabilistic model
  for code with decision trees}.
\newblock In \emph{Proceedings of the 2016 {ACM} {SIGPLAN} International
  Conference on Object-Oriented Programming, Systems, Languages, and
  Applications, {OOPSLA} 2016, part of {SPLASH} 2016, Amsterdam, The
  Netherlands, October 30 - November 4, 2016}, pages 731--747. {ACM}.

\bibitem[{Sennrich et~al.(2016)Sennrich, Haddow, and
  Birch}]{Sennrich2016NeuralMT}
Rico Sennrich, Barry Haddow, and Alexandra Birch. 2016.
\newblock Neural machine translation of rare words with subword units.
\newblock \emph{ArXiv}, abs/1508.07909.

\bibitem[{Svyatkovskiy et~al.(2020)Svyatkovskiy, Deng, Fu, and
  Sundaresan}]{svyatkovskiy2020intellicode}
Alexey Svyatkovskiy, Shao~Kun Deng, Shengyu Fu, and Neel Sundaresan. 2020.
\newblock \href {https://doi.org/10.1145/3368089.3417058} {Intellicode compose:
  code generation using transformer}.
\newblock In \emph{{ESEC/FSE} '20: 28th {ACM} Joint European Software
  Engineering Conference and Symposium on the Foundations of Software
  Engineering, Virtual Event, USA, November 8-13, 2020}, pages 1433--1443.
  {ACM}.

\bibitem[{Vaswani et~al.(2017)Vaswani, Shazeer, Parmar, Uszkoreit, Jones,
  Gomez, Kaiser, and Polosukhin}]{vaswani2017attention}
Ashish Vaswani, Noam Shazeer, Niki Parmar, Jakob Uszkoreit, Llion Jones,
  Aidan~N. Gomez, Lukasz Kaiser, and Illia Polosukhin. 2017.
\newblock \href
  {https://proceedings.neurips.cc/paper/2017/hash/3f5ee243547dee91fbd053c1c4a845aa-Abstract.html}
  {Attention is all you need}.
\newblock In \emph{Advances in Neural Information Processing Systems 30: Annual
  Conference on Neural Information Processing Systems 2017, December 4-9, 2017,
  Long Beach, CA, {USA}}, pages 5998--6008.

\bibitem[{Wan et~al.(2022{\natexlab{a}})Wan, He, Bi, Zhang, Sui, Zhang,
  Hashimoto, Jin, Xu, Xiong, and Yu}]{wan2022naturalcc}
Yao Wan, Yang He, Zhangqian Bi, Jianguo Zhang, Yulei Sui, Hongyu Zhang, Kazuma
  Hashimoto, Hai Jin, Guandong Xu, Caiming Xiong, and Philip~S. Yu.
  2022{\natexlab{a}}.
\newblock Naturalcc: An open-source toolkit for code intelligence.
\newblock In \emph{Proceedings of 44th International Conference on Software
  Engineering, Companion Volume}. {ACM}.

\bibitem[{Wan et~al.(2019)Wan, Shu, Sui, Xu, Zhao, Wu, and
  Yu}]{Wan2019MultimodalAN}
Yao Wan, Jingdong Shu, Yulei Sui, Guandong Xu, Zhou Zhao, Jian Wu, and
  Philip~S. Yu. 2019.
\newblock \href {https://doi.org/10.1109/ASE.2019.00012} {Multi-modal attention
  network learning for semantic source code retrieval}.
\newblock In \emph{34th {IEEE/ACM} International Conference on Automated
  Software Engineering, {ASE} 2019, San Diego, CA, USA, November 11-15, 2019},
  pages 13--25. {IEEE}.

\bibitem[{Wan et~al.(2022{\natexlab{b}})Wan, Zhao, Zhang, Sui, Xu, and
  Jin}]{wan2022they}
Yao Wan, Wei Zhao, Hongyu Zhang, Yulei Sui, Guandong Xu, and Hai Jin.
  2022{\natexlab{b}}.
\newblock What do they capture?--a structural analysis of pre-trained language
  models for source code.
\newblock In \emph{Proceedings of the 44th International Conference on Software
  Engineering}.

\bibitem[{Wan et~al.(2018)Wan, Zhao, Yang, Xu, Ying, Wu, and
  Yu}]{wan2018improving}
Yao Wan, Zhou Zhao, Min Yang, Guandong Xu, Haochao Ying, Jian Wu, and Philip~S
  Yu. 2018.
\newblock Improving automatic source code summarization via deep reinforcement
  learning.
\newblock In \emph{Proceedings of the 33rd ACM/IEEE International Conference on
  Automated Software Engineering}, pages 397--407.

\bibitem[{Wang and Su(2020)}]{Wang2020BlendedPS}
Ke~Wang and Zhendong Su. 2020.
\newblock \href {https://doi.org/10.1145/3385412.3385999} {Blended, precise
  semantic program embeddings}.
\newblock In \emph{Proceedings of the 41st {ACM} {SIGPLAN} International
  Conference on Programming Language Design and Implementation, {PLDI} 2020,
  London, UK, June 15-20, 2020}, pages 121--134. {ACM}.

\bibitem[{Wang et~al.(2020{\natexlab{a}})Wang, Liu, Li, Chen, Liu, and
  Wu}]{Wang20ASE}
Xin Wang, Jin Liu, Li~Li, Xiao Chen, Xiao Liu, and Hao Wu. 2020{\natexlab{a}}.
\newblock \href {https://doi.org/10.1145/3324884.3416583} {Detecting and
  explaining self-admitted technical debts with attention-based neural
  networks}.
\newblock In \emph{35th {IEEE/ACM} International Conference on Automated
  Software Engineering, {ASE} 2020, Melbourne, Australia, September 21-25,
  2020}, pages 871--882. {IEEE}.

\bibitem[{Wang et~al.(2020{\natexlab{b}})Wang, Liu, Liu, Cui, and
  Wu}]{Wang20SS}
Xin Wang, Jin Liu, Xiao Liu, Xiaohui Cui, and Hao Wu. 2020{\natexlab{b}}.
\newblock \href {https://doi.org/10.1007/978-3-030-60259-8\_56} {A spatial and
  sequential combined method for web service classification}.
\newblock In \emph{Web and Big Data - 4th International Joint Conference,
  APWeb-WAIM 2020, Tianjin, China, September 18-20, 2020, Proceedings, Part
  {I}}, volume 12317 of \emph{Lecture Notes in Computer Science}, pages
  764--778. Springer.

\bibitem[{Wang et~al.(2021{\natexlab{a}})Wang, Liu, Li, Chen, Liu, and
  Wu}]{Wangicwspoi}
Xin Wang, Xiao Liu, Li~Li, Xiao Chen, Jin Liu, and Hao Wu. 2021{\natexlab{a}}.
\newblock \href {https://doi.org/10.1109/ICWS53863.2021.00028} {Time-aware user
  modeling with check-in time prediction for next {POI} recommendation}.
\newblock In \emph{2021 {IEEE} International Conference on Web Services, {ICWS}
  2021, Chicago, IL, USA, September 5-10, 2021}, pages 125--134. {IEEE}.

\bibitem[{Wang et~al.(2021{\natexlab{b}})Wang, Liu, Liu, and Wu}]{Wangbundle21}
Xin Wang, Xiao Liu, Jin Liu, and Hao Wu. 2021{\natexlab{b}}.
\newblock \href {https://doi.org/10.1109/ICWS53863.2021.00033} {Relational
  graph neural network with neighbor interactions for bundle recommendation
  service}.
\newblock In \emph{2021 {IEEE} International Conference on Web Services, {ICWS}
  2021, Chicago, IL, USA, September 5-10, 2021}, pages 167--172. {IEEE}.

\bibitem[{{Wang} et~al.(2021){Wang}, {Wang}, {Mi}, {Zhou}, {Wan}, {Liu}, {Li},
  {Wu}, {Liu}, and {Jiang}}]{Wang2021SynCoBERTSM}
Xin {Wang}, Yasheng {Wang}, Fei {Mi}, Pingyi {Zhou}, Yao {Wan}, Xiao {Liu},
  Li~{Li}, Hao {Wu}, Jin {Liu}, and Xin {Jiang}. 2021.
\newblock Syncobert: Syntax-guided multi-modal contrastive pre-training for
  code representation.

\bibitem[{Wang et~al.(2022)Wang, Wang, Wan, Mi, Li, Zhou, Liu, Wu, Jiang, and
  Liu}]{wang2022}
Xin Wang, Yasheng Wang, Yao Wan, Fei Mi, Yitong Li, Pingyi Zhou, Jin Liu, Hao
  Wu, Xin Jiang, and Qun Liu. 2022.
\newblock \href {https://doi.org/10.48550/arXiv.2203.05132} {Compilable neural
  code generation with compiler feedback}.
\newblock volume abs/2203.05132.

\bibitem[{Wang et~al.(2021{\natexlab{a}})Wang, Zhou, Wang, Liu, Liu, and
  Wu}]{Wang21ICSOC}
Xin Wang, Pingyi Zhou, Yasheng Wang, Xiao Liu, Jin Liu, and Hao Wu.
  2021{\natexlab{a}}.
\newblock \href {https://doi.org/10.1007/978-3-030-91431-8\_29} {Servicebert:
  {A} pre-trained model for web service tagging and recommendation}.
\newblock In \emph{Service-Oriented Computing - 19th International Conference,
  {ICSOC} 2021, Virtual Event, November 22-25, 2021, Proceedings}, volume 13121
  of \emph{Lecture Notes in Computer Science}, pages 464--478. Springer.

\bibitem[{Wang et~al.(2021{\natexlab{b}})Wang, Wang, Joty, and
  Hoi}]{Wang2021CodeT5IU}
Yue Wang, Weishi Wang, Shafiq~R. Joty, and Steven C.~H. Hoi.
  2021{\natexlab{b}}.
\newblock \href {https://doi.org/10.18653/v1/2021.emnlp-main.685} {Codet5:
  Identifier-aware unified pre-trained encoder-decoder models for code
  understanding and generation}.
\newblock In \emph{Proceedings of the 2021 Conference on Empirical Methods in
  Natural Language Processing, {EMNLP} 2021, Virtual Event / Punta Cana,
  Dominican Republic, 7-11 November, 2021}, pages 8696--8708. Association for
  Computational Linguistics.

\bibitem[{White et~al.(2016)White, Tufano, Vendome, and
  Poshyvanyk}]{White2016DeepLC}
Martin White, Michele Tufano, Christopher Vendome, and Denys Poshyvanyk. 2016.
\newblock \href {https://doi.org/10.1145/2970276.2970326} {Deep learning code
  fragments for code clone detection}.
\newblock In \emph{Proceedings of the 31st {IEEE/ACM} International Conference
  on Automated Software Engineering, {ASE} 2016, Singapore, September 3-7,
  2016}, pages 87--98. {ACM}.

\bibitem[{Wu et~al.(2021)Wu, Duan, Yue, and Zhang}]{wu2021}
Hao Wu, Yunhao Duan, Kun Yue, and Lei Zhang. 2021.
\newblock \href {https://doi.org/10.1109/TSC.2021.3098756} {Mashup-oriented web
  api recommendation via multi-model fusion and multi-task learning}.
\newblock \emph{IEEE Transactions on Services Computing}, pages 1--1.

\bibitem[{Yin et~al.(2018)Yin, Deng, Chen, Vasilescu, and
  Neubig}]{Yin2018LearningTM}
Pengcheng Yin, Bowen Deng, Edgar Chen, Bogdan Vasilescu, and Graham Neubig.
  2018.
\newblock \href {https://doi.org/10.1145/3196398.3196408} {Learning to mine
  aligned code and natural language pairs from stack overflow}.
\newblock In \emph{Proceedings of the 15th International Conference on Mining
  Software Repositories, {MSR} 2018, Gothenburg, Sweden, May 28-29, 2018},
  pages 476--486. {ACM}.

\bibitem[{Yu et~al.(2022)Yu, Zhao, Liu, Liu, Xu, and Wang}]{YuZLLXW22}
Jiaojiao Yu, Kunsong Zhao, Jin Liu, Xiao Liu, Zhou Xu, and Xin Wang. 2022.
\newblock \href {https://doi.org/10.1016/j.jss.2022.111219} {Exploiting gated
  graph neural network for detecting and explaining self-admitted technical
  debts}.
\newblock \emph{J. Syst. Softw.}, 187:111219.

\bibitem[{Zhao et~al.(2022)Zhao, Liu, Xu, Liu, Xue, Xie, Zhou, and
  Wang}]{ZHAO2022111324}
Kunsong Zhao, Jin Liu, Zhou Xu, Xiao Liu, Lei Xue, Zhiwen Xie, Yuxuan Zhou, and
  Xin Wang. 2022.
\newblock \href {https://doi.org/https://doi.org/10.1016/j.jss.2022.111324}
  {Graph4web: A relation-aware graph attention network for web service
  classification}.
\newblock \emph{Journal of Systems and Software}, 190:111324.

\bibitem[{Zhao et~al.(2021{\natexlab{a}})Zhao, Xu, Yan, Zhang, Yang, and
  Li}]{Zhao0Y00021}
Kunsong Zhao, Zhou Xu, Meng Yan, Tao Zhang, Dan Yang, and Wei Li.
  2021{\natexlab{a}}.
\newblock \href {https://doi.org/10.1016/j.infsof.2021.106652} {A comprehensive
  investigation of the impact of feature selection techniques on crashing fault
  residence prediction models}.
\newblock \emph{Inf. Softw. Technol.}, 139:106652.

\bibitem[{Zhao et~al.(2021{\natexlab{b}})Zhao, Xu, Zhang, Tang, and
  Yan}]{ZhaoXZTY21}
Kunsong Zhao, Zhou Xu, Tao Zhang, Yutian Tang, and Meng Yan.
  2021{\natexlab{b}}.
\newblock \href {https://doi.org/10.1109/TR.2021.3060937} {Simplified deep
  forest model based just-in-time defect prediction for android mobile apps}.
\newblock \emph{{IEEE} Trans. Reliab.}, 70(2):848--859.

\end{thebibliography}
\bibliographystyle{acl_natbib}




\end{document}